\documentclass{article} 
\usepackage{iclr2021_conference,times}

\usepackage{amsmath,amsfonts,bm}

\def\eqref#1{equation~\ref{#1}}

\def\1{\bm{1}}

\DeclareMathAlphabet{\mathsfit}{\encodingdefault}{\sfdefault}{m}{sl}
\SetMathAlphabet{\mathsfit}{bold}{\encodingdefault}{\sfdefault}{bx}{n}

\usepackage{hyperref}
\usepackage{graphicx}
\usepackage{url}

\title{MIT SafePaths Card (MiSaCa): Augmenting Paper Based Vaccination Cards with Printed Codes}

\author{\\\textbf{Joseph Bae}\textsuperscript{1, 3}, \textbf{Rohan Sukumaran}\textsuperscript{1}, \textbf{Sheshank Shankar}\textsuperscript{1}, \textbf{Saurish Srivastava}\textsuperscript{1}, \textbf{Rohan Iyer}\textsuperscript{1}, \\\textbf{Aryan Mahindra}\textsuperscript{1}, \textbf{Qamil Mirza}\textsuperscript{1},  \textbf{Maurizio Arseni}\textsuperscript{1},  \textbf{Anshuman Sharma}\textsuperscript{1},  \textbf{Saras Agrawal}\textsuperscript{1}, \\ \textbf{Orna}  \textbf{Mukhopadhyay}\textsuperscript{1}, \textbf{Colin Kang}\textsuperscript{1},  \textbf{Priyanshi Katiyar}\textsuperscript{1},  \textbf{Apurv Shekhar}\textsuperscript{1},  \textbf{Sifat Hasan}\textsuperscript{1},\\ \textbf{Krishnendu} \textbf{Dasgupta}\textsuperscript{1}, \textbf{Darshan Gandhi}\textsuperscript{1},  \textbf{Sethuraman TV}\textsuperscript{1},  \textbf{Parth Patwa}\textsuperscript{1},  \textbf{Ishaan Singh}\textsuperscript{1},\\ \textbf{Abhishek Singh}\textsuperscript{2}, \textbf{Ramesh Raskar}\textsuperscript{1,2}  \\\\
\textsuperscript{1}PathCheck Foundation, 02139 Cambridge, USA.\\
\textsuperscript{2}MIT Media Lab, 02139 Cambridge, USA.\\
\textsuperscript{3}Renaissance School of Medicine, Stony Brook University, 11794 Stony Brook, USA.\\
\texttt{raskar@media.mit.edu} }

\iclrfinalcopy 
\begin{document}

\maketitle

\begin{abstract}
    In this early draft,  we describe a user-centric, card-based system for vaccine distribution. Our system makes use of digitally signed QR codes and their use for phased vaccine distribution, vaccine administration/record-keeping, immunization verification, and follow-up symptom reporting. Furthermore, we propose and describe a complementary scanner app system to be used by vaccination clinics, public health officials, and immunization verification parties to effectively utilize card-based framework. We believe that the proposed system provides a privacy-preserving and efficient framework for vaccine distribution in both developed and developing regions. 
\end{abstract}

\section{Introduction}
Without an effective curative or preventative measure, the unprecedented coronavirus disease 2019 (COVID-19) pandemic has led to a significant amount of human deaths (1,900,000 at the time of publication (\cite{webtrack})). However, now with the advent of vaccines, we face the challenges of strategic, equitable and privacy preserved ways for last-mile vaccine distribution (\cite{bae2020challenges, vaccinetrack}).   

First, the vaccine recipients must be dynamically prioritized to ensure an equitable reach, especially as multiple vaccines with different protocols are approved in various areas. In addition, once a citizen’s first dose is administered, they must follow through with their second dose as well. Also, a communication plan must also be put in place to combat inevitable rumours, misinformation, and conspiracy theories aiming to disrupt citizen engagement in the vaccination process (~\cite{morales2021covid19, wp_article}). It must also address the mistrust of vaccines in society (\cite{Palamenghi2020}), especially within previously marginalized minority populations (\cite{JHU}). This is why we must take a user-centric approach that preserves trust — vaccines are meaningless if citizens aren’t willing to take them (\cite{PMC}). Lastly, the health outcomes (effectiveness, safety, long-term effects, etc) of the vaccines must be effectively monitored in a privacy-preserving way (\cite{hbr}).

In today’s society, multiple technological systems are being utilized by the Center for Disease Control (CDC) to combat these challenges (~\cite{cdc1,cdc2,cdc3}). For example, the Vaccine Administration Management System (VAMS) streamlines the vaccine distribution process for jurisdictions, employers, and healthcare providers. In addition, it’s an effective user-centric system as it allows for vaccine recipients to schedule appointments, receive records of their visit, and receive reminders for a second dose (\cite{VAMS}). The Immunization Information Systems (IIS) are a group of privacy preserving database systems that track all vaccinations within various areas (\cite{IIS}). Lastly, the Vaccine Adverse Event Reporting System (VAERS) is the prominent system for the monitoring of health outcomes (~\cite{VAERS,vaers_article}).

In our previous work, we detail the MIT SafePaths app-based protocol for vaccine distribution. In this paper, we introduce a separate user-centric card protocol that uses printed codes as a supplement to traditional paper based vaccination cards. 
\section{Card Flow}
\subsection{Overview}
Here we present a vaccine distribution system utilizing physical SafePaths cards and four digitally signed QR code stickers (henceforth termed \textit{Coupon}, \textit{Badge}, \textit{Passkey}, and \textit{Status}). The digital signing of a QR code is simply a secure process of verifying the authenticity of the information contained in the QR code (\cite{crypt}). These QR code stickers are simply QR codes printed onto adhesive stickers that can then be attached to a user’s physical card. 

\begin{figure}[ht!]
\begin{center}
\includegraphics[width=12cm]{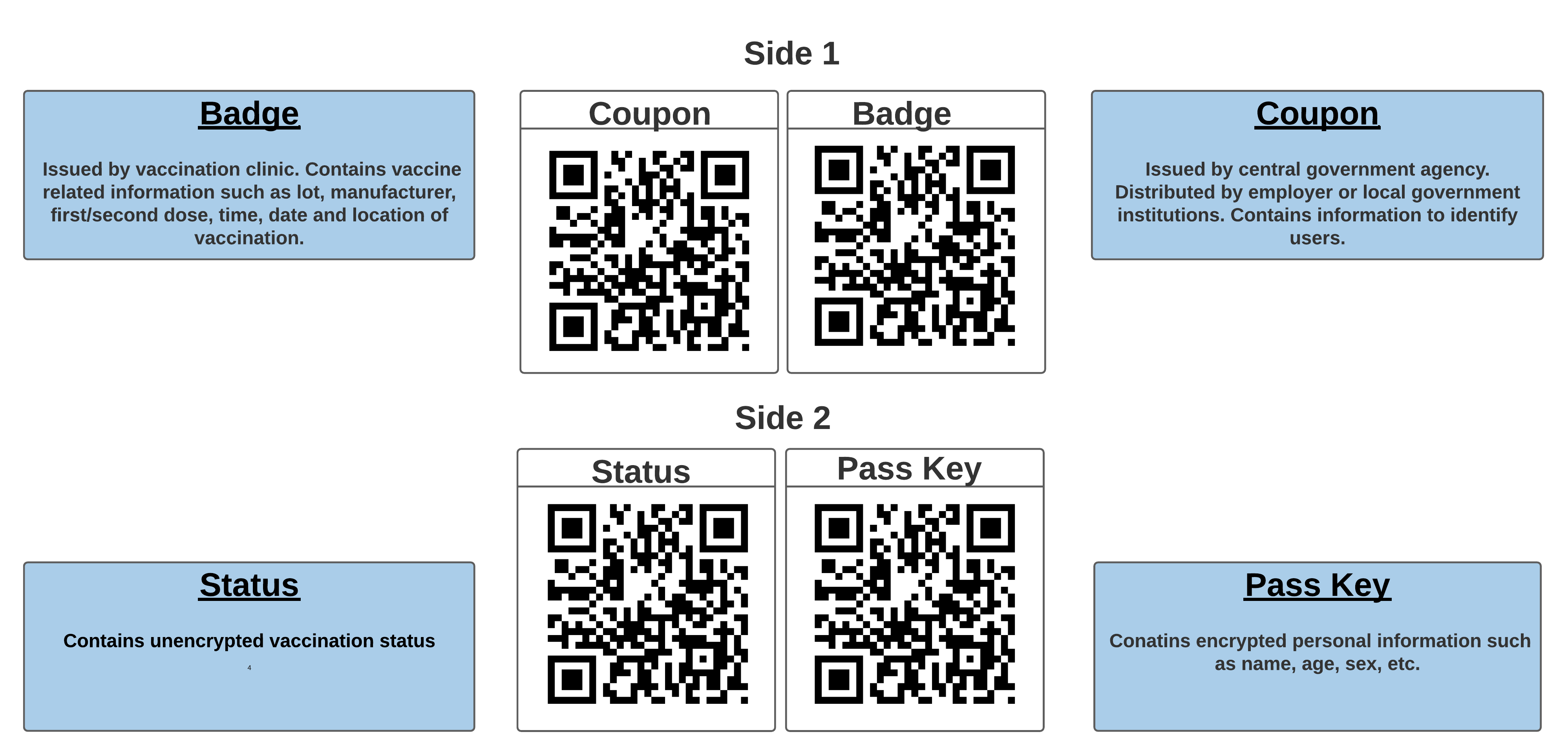}
\end{center}
\caption{The 4 digitally signed QR code stickers (\textit{Coupon}, \textit{Badge}, \textit{Status}, and \textit{Passkey}) present on the SafePaths cards.}
\label{qr-codes}
\end{figure}

\begin{figure}[ht!]
\begin{center}
\includegraphics[width=13.5cm]{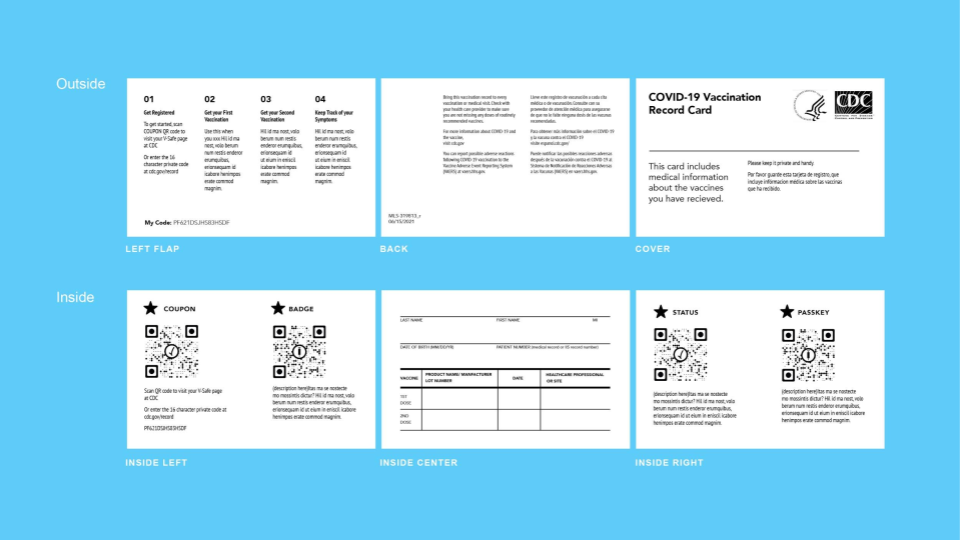}
\end{center}
\caption{MIT SafePaths Card Mockup}
\label{mockup}
\end{figure}
The digital signature of the QR codes take place as below
\textbf{Certificate = (message, signature(messages))}

For each sticker below the message is as follows - 
\begin{itemize}
    \item \textbf{Coupon} = (number, total, city, phase, (age, job, comorbidities/sick))  
    \item \textbf{Badge} = (coupon, dose\_info, Hash(passkey())
    \item \textbf{Status} = ((vaccinated = 0,1,2), Hash(passkey())
    \item \textbf{Passkey} = (name, DOB, salt) \\
    = hash:sj2d8k8hy7j
    
\end{itemize}
\begin{table}[!htb]
\begin{tabular}{|l|l|l|l|}
\hline
\multicolumn{1}{|c|}{\textbf{Name}} &
  \multicolumn{1}{c|}{\textbf{Equation}} &
  \multicolumn{1}{c|}{\textbf{Description}} &
  \multicolumn{1}{c|}{\textbf{Example}} \\ \hline
Coupon &
  \begin{tabular}[c]{@{}l@{}}\{m, sign(m)\} \\ where m = (i, zip code, \\ job type)\end{tabular} &
  \begin{tabular}[c]{@{}l@{}}Coupon code is signed \\ by CDC and indicates the\\  zip code and job type of the \\ receiver\end{tabular} &
  \begin{tabular}[c]{@{}l@{}}\{37, 5000, Springfield, 1B, \\ Teacher\}\end{tabular} \\ \hline
Badge &
  \begin{tabular}[c]{@{}l@{}}\{m, sign(m)\} \\ where m = (dose\_info, \\ coupon, hash(passkey))\end{tabular} &
  \begin{tabular}[c]{@{}l@{}}Badge is available after 2 \\ doses and it gives the \\ information pertinent to \\ the vaccine shot.\end{tabular} &
  \begin{tabular}[c]{@{}l@{}}\{{[}Pfizer, “1st Dose”, \\ 1/1/2021{]}, fe4c2, \\ 3be33c20cc4c85a0c32f7bf5b4\}\end{tabular} \\ \hline
Status &
  \begin{tabular}[c]{@{}l@{}}\{m, sign(m)\} \\ where m = (status, \\ hash(passkey))\end{tabular} &
  \begin{tabular}[c]{@{}l@{}}Contains bare minimum \\ information to prove the \\ user is vaccinated\end{tabular} &
  \begin{tabular}[c]{@{}l@{}}\{vaccinated, \\ 3be33c20cc4c85a0c32f7bf5b4\}\end{tabular} \\ \hline
Passkey &
  \begin{tabular}[c]{@{}l@{}}ID = User\_PII\\ Key = salt\end{tabular} &
  \begin{tabular}[c]{@{}l@{}}Key is the random salt \\ used for increasing the \\ entropy of hashed data\end{tabular} &
  \begin{tabular}[c]{@{}l@{}}\{John Doe, \\ 6363fe744f74ee8f280958\}\end{tabular} \\ \hline
\end{tabular}
\caption{This outlines the four QR codes and what information is digitally signed in it.}
\label{tab:my-table}
\end{table}

Our solution is intended to decouple the health information and personally identifiable information (PII) in this process. Thereby, we are essentially proposing to separate the eligibility of the vaccination from the distribution of it. This way we can have the health information centralised, whilst the PII information decentralised.

\subsection{Vaccine eligibility confirmation}
To accommodate the several-stage vaccination policies that countries have begun to employ, SafePaths cards will be distributed containing one digitally-signed \textit{Coupon} QR code. This would be provided by a central government agency such as the CDC and made available to users either by an employer or local government location. A pseudo random identifier generated for this \textit{Coupon} serves as the identifying information for the user throughout the remaining workflow. This \textit{Coupon} would initially come with  SafePaths cards while the remaining three adhesive stickers must be obtained and placed onto the card following vaccination events.

\subsection{Vaccine administration}
Check-in at a vaccination clinic would require the verification  of a user’s \textit{Coupon}. 

Upon vaccination, the vaccination clinic would create a digitally-signed record of immunization and print it as a QR code on an adhesive sticker. This adhesive sticker (henceforth referred to as the \textit{Badge}) would contain information regarding vaccine lot, manufacturer, and first/second dose information. The \textit{Badge} would also contain information regarding the time, date, and location of vaccination. 

The vaccination clinic would also create a unique encryption key to encrypt the \textit{Badge}. This key, as well as encrypted PII such as name, age, sex, etc. would be stored on a \textit{Passkey} QR code, printed onto a \textit{Passkey} QR sticker. This \textit{Passkey} is required for decryption of PII and in-depth vaccination information (time, date, location of vaccination). 

At this stage, a vaccine recipient would then have \textit{Coupon}, \textit{Badge}, and \textit{Passkey} QR stickers.

\subsection{Second Dose}
When a user attempts to receive a second dose of a vaccine, the vaccination clinic would  utilize a user’s \textit{Badge} to determine the appropriate vaccine type and dose and the \textit{Passkey} to confirm a user’s identity. Again, the user \textit{Passkey} contains information that solely exists on the physical card carried by a user. Use of this sticker is required to decrypt in depth vaccination information for a patient contained in the \textit{Badge} (location of vaccination, date, etc.). Once final vaccination has been performed, the vaccine clinic would create a fourth and final \textit{Status} QR code sticker for a recipient’s SafePaths card, which would simply indicate whether or not a user has been vaccinated. \textit{Status} would not contain any further information and therefore would be unencrypted.  
\subsection{Record-keeping}
User vaccination records could be linked by anonymized upload to a centralized system using a user’s pseudorandom identifier. The user’s \textit{Passkey}, containing their encryption key that decrypts their PII, would not be uploaded to the CDC without consent. Alternatively, we propose an anonymous record keeping function in our Scanner App section. 
\subsection{Vaccination verification}
Verification of immunization status might be required in various scenarios such as airline travel, return to school/work, etc. Vaccine verification at these venues would follow the receipt of a second COVID-19 dose. 

Information regarding an individual’s vaccination status would be digitally signed by the vaccine clinic onto the \textit{Status} sticker. When scanned, this sticker would provide the verifier with information regarding whether or not an individual has been vaccinated. If further verification of identity is required, the verifier could make use of a consenting individual’s \textit{Passkey} sticker to decrypt the holder’s name. With this method, a user would have multiple levels of information they can share, beginning with vaccination status in the unencrypted \textit{Status} sticker, basic personal information (i.e. name) that must be decrypted using the \textit{Passkey} sticker, and finally full personal vaccination information encrypted in the \textit{Badge}.  

\subsection{Safety and efficacy monitoring}
Short and long-term monitoring of health outcomes would rely on self-reporting. These cards could still facilitate the anonymous information upload by interacting with existing centralized systems such as VAERS or V-Safe while bypassing PII input. All health and symptom information could instead be tied to a user’s pseudorandom ID. We also propose a scanner app solution in the Scanner Flow section that could aggregate symptom reporting and vaccine record data anonymously.  

\begin{figure}[ht!]
\begin{center}
\includegraphics[width=13.5cm]{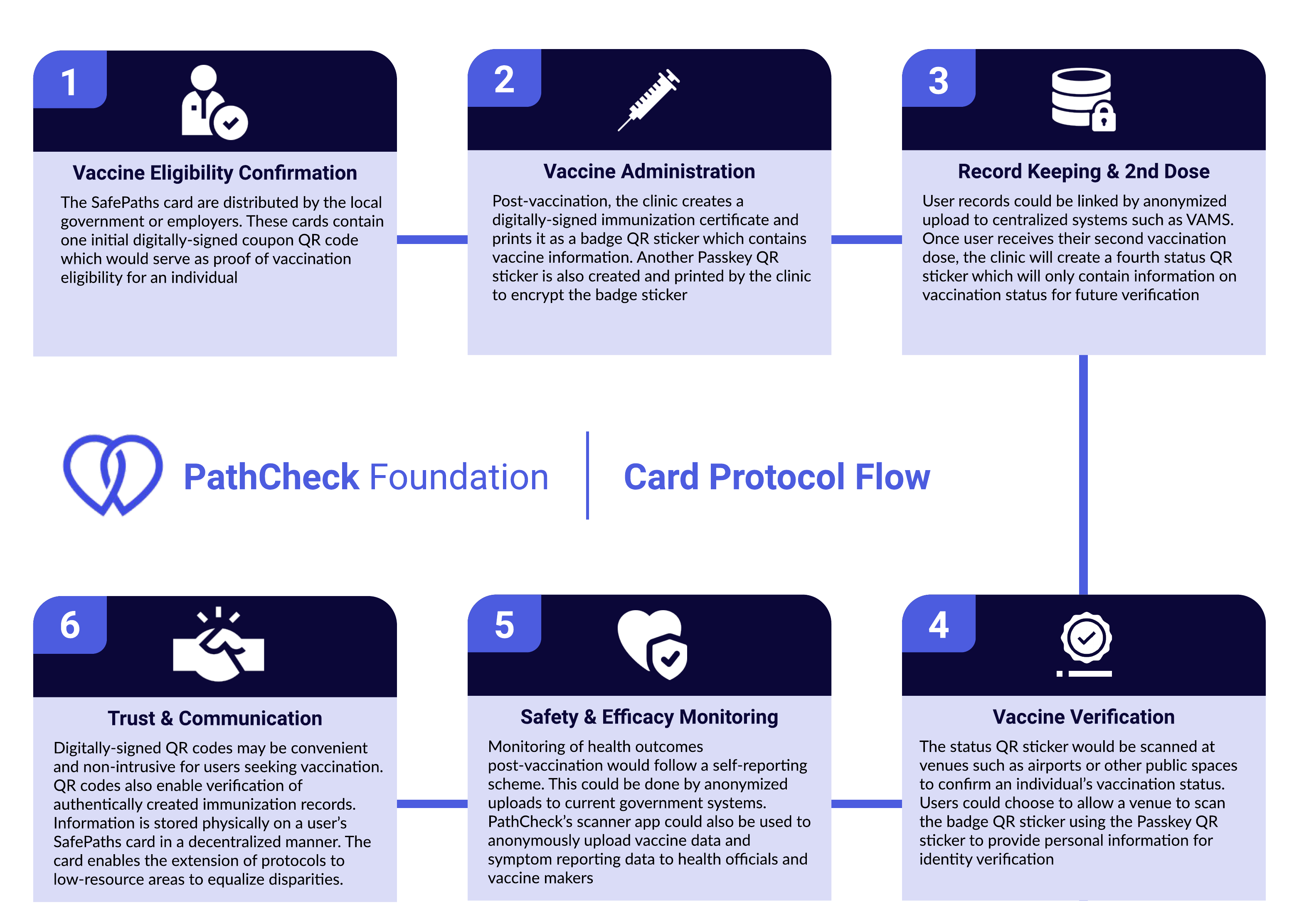}
\end{center}
\caption{Card protocol workflow diagram.}
\label{card-protocol}
\end{figure}
\section{Scanner Flow}

\subsection{Overview}
Here we discuss the systems that must be built for vaccine clinics and distributors in order to enable the use of the SafePaths card framework presented above. We present several relevant protocols as well as the functionality of a proposed vaccine distributor/verifier scanner app. This scanner app would be necessary to function with the encrypted QR codes described above. 

\subsection{Vaccine eligibility confirmation}
Phased vaccination using the SafePaths card system requires the distribution of SafePaths cards containing digitally signed \textit{Coupons} to appropriate subsets of the population during each stage of vaccination. There are several ways that this might be achieved. We propose potential solutions below, though we recognize that these strategies must be determined by individual jurisdictions to meet the circumstances in different locations.

\begin{enumerate}
    \item Disseminate to businesses to provide to employees (eg: hospitals, restaurants, etc. as appropriate)
    \item Make available at local government building (similar to DMV process of obtaining a driver’s license)
    \item Mail out to individuals based on employment/other factors (via background check systems, centralized databases such as IRS) 
\end{enumerate}

\begin{figure}[!ht]
\begin{center}
\includegraphics[width=13.5cm]{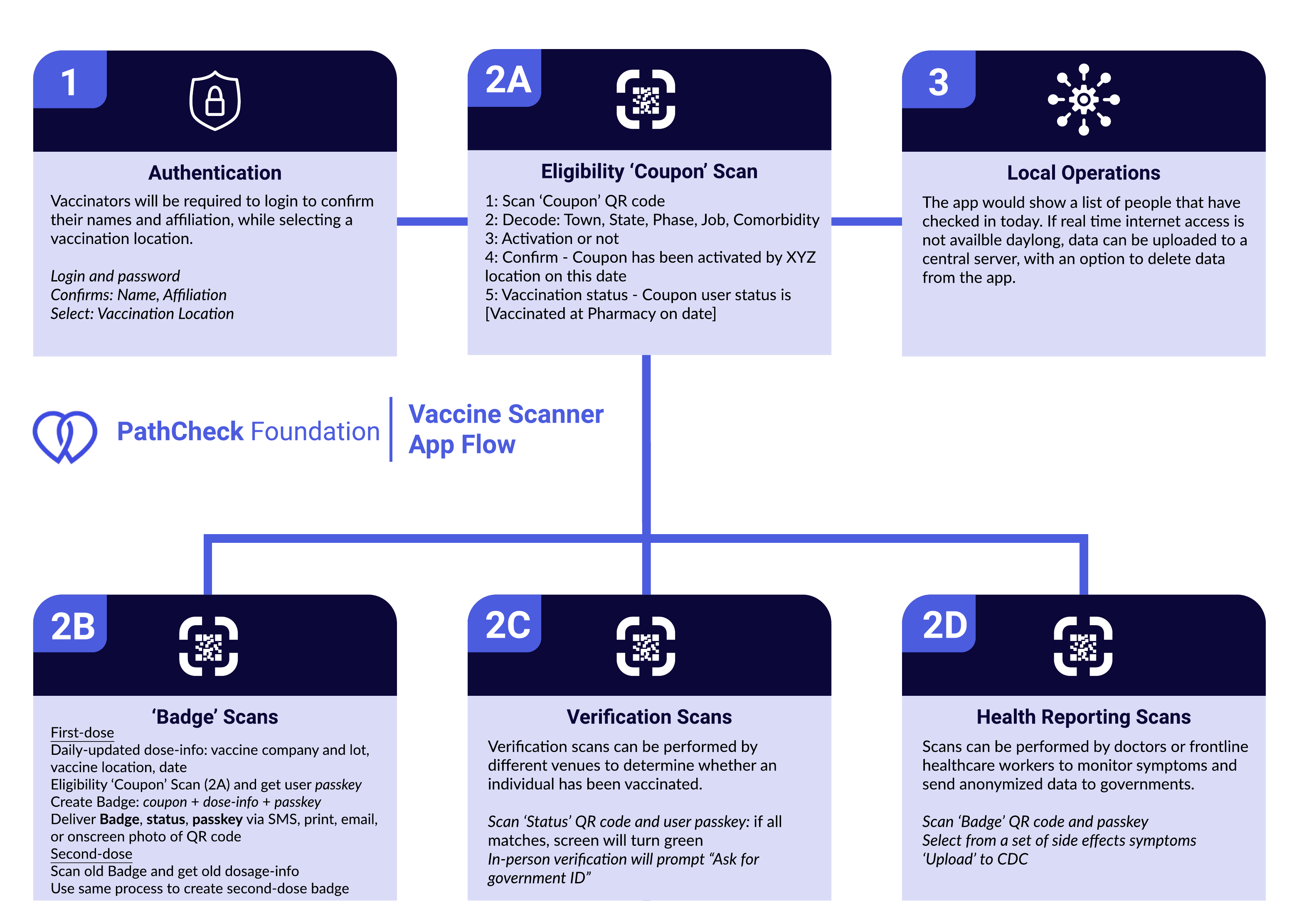}
\end{center}
\caption{Scanner app protocol workflow diagram.}
\label{scanner-workflow}
\end{figure}

\subsection{Vaccine administration}
To confirm an individual for vaccination scheduling/check-in, a clinic must verify the authenticity of a vaccine recipient’s QR \textit{Coupons}. The first function of our proposed scanner app would be to scan a vaccine recipient’s \textit{Coupon} to determine authenticity and prevent the use of a single \textit{Coupon} by multiple individuals. This would be achieved by scanning the digital signature present on a SafePaths \textit{Coupon} and verifying its digital signature. 

The second function of our proposed scanner app would be to create digitally signed \textit{Badge} and \textit{Passkey} stickers for post vaccination. This would make use of our previously described algorithm (\cite{crypt}) for secure recording of vaccine information into a \textit{Badge} sticker, encrypted using the encryption key present in the \textit{Passkey}. After creating these stickers, the proposed scanner app would not store any information regarding a recipient’s encryption key; that information would only exist within the \textit{Passkey} sticker. 

\subsection{Second dose}

Second dose administration functionality would be implemented into the scanner app in the same manner as described in the previous section for ‘Vaccine Administration’. A \textit{Status} sticker would be created by the scanner app in a similar manner to the \textit{Badge} sticker, also drawing on the methods described in our cryptographic protocol (\cite{crypt}). 

\subsection{Record-keeping}
Another critical function of our scanner app would be the ability to integrate with existing systems, such as VAMS in the United States. Ideally, our app would be able to automatically provide vaccination record information to VAMS while replacing PII with pseudo identifiers. 

Alternatively, our scanner system would also have the capability to directly aggregate vaccination record data in an anonymized fashion, retaining population-level statistics such as vaccination prevalence in a given jurisdiction that might be important for public health policy development. Details concerning clinic location, vaccine dose, and vaccine manufacturer could be stored by the scanner app and aggregated for public health official viewing.

\subsection{Vaccination verification}
Our proposed scanner app would enable vaccination verification simply by reading immunization status contained in a user’s \textit{Status} sticker. For further identity verification, a form of ID (such as driver’s license) can be compared with the decrypted PII from the scanner app using an individual’s \textit{Passkey} sticker.  The scanner app would not store this information following completion of the immunization confirmation.

\section{Conclusion}
In this early draft, we present a complete protocol for a physical card-based system for phased vaccine distribution, individual vaccination, second-dose adherence, and symptom follow-up. Due to their physical nature and simplicity, digitally-signed QR codes may be a convenient and non-intrusive modality for some users seeking vaccination. Digitally-signed QR stickers enable verification of authentically created immunization records, and the encryption schema presented using a unique passkey sticker ensures that user PII can only be decrypted with the user’s consent. This information is stored physically on the user’s SafePaths card in a decentralized manner wherein a user must provide their physical passkey sticker for decryption of PII. These cards also extend privacy-focused protocols to low-resource areas and populations, equalizing disparities in access to individual-centric solutions and frameworks for COVID-19 vaccination. The centralised health data collected (which is rid of all PIIs) can be used by the concerned authorities to have population aggregated view of the vaccine adherence in a region. Furthermore, such privacy preserving dashboards which show aggregated data can help the authorities take informed decisions.

\subsubsection*{Acknowledgments}
We are grateful to Riyanka Roy Choudhury, CodeX Fellow, Stanford University, Adam Berrey, CEO of PathCheck Foundation, Dr. Brooke Struck, Research Director at The Decision Lab, Canada, Vinay Gidwaney, Entrepreneur and Advisor, PathCheck Foundation, and Paola Heudebert, co-founder of Blockchain for Human Rights, Alison Tinker, Saswati Soumya, Sunny Manduva, Bhavya Pandey, and Aarathi Prasad for their assistance in discussions, support and guidance in writing of this paper.

\bibliography{iclr2021_conference}
\bibliographystyle{iclr2021_conference}

\end{document}